\documentclass[12pt]{article}
\usepackage{cite} %use drftcite in draftmode otherwise
\usepackage{epsf}
\usepackage{pstricks}
\usepackage{textcomp}
\usepackage{epsf,epsfig,subfigure,graphicx,amsmath,amssymb}
\usepackage{color}

\def\lsim{\mathrel{\rlap{\lower4pt\hbox{\hskip1pt$\sim$}}
    \raise1pt\hbox{$<$}}}         %less than or approx. symbol
\def\gsim{\mathrel{\rlap{\lower4pt\hbox{\hskip1pt$\sim$}}
    \raise1pt\hbox{$>$}}}         %greater than or approx. symbol\newcommand{\beq}{\begin{eqnarray}}% can be used as {equation} or {eqnarray}
\newcommand{\eeq}{\end{eqnarray}}

\newcommand{\centeron}[2]{{\setbox0=\hbox{#1}\setbox1=\hbox{#2}\ifdim
                           \wd1>\wd0\kern.5\wd1\kern-.5\wd0\fi \copy0
                           \kern-.5\wd0\kern-.5\wd1\copy1\ifdim\wd0>\wd1
                           \kern.5\wd0\kern-.5\wd1\fi}}
\newcommand{\ltap}{\>\centeron{\raise.35ex\hbox{$<$}}
                   {\lower.65ex\hbox{$\sim$}}\>}
\newcommand{\gtap}{\>\centeron{\raise.35ex\hbox{$>$}}
                   {\lower.65ex\hbox{$\sim$}}\>}

\newcommand\ZZ{\hbox{\zfont Z\kern-.4emZ}}
\font\zfont = cmss10 %scaled \magstep1

\newcommand{\be}{\begin{equation}}
\newcommand{\ee}{\end{equation}}
\newcommand{\bea}{\begin{eqnarray}}
\newcommand{\eea}{\end{eqnarray}}

\newcommand{\beq}{\begin{equation}}% can be used as {equation} 

\begin{document}

%%%%%%%%%%%%%%%%%%%%%%%%%%%%%%%%%%%%%%%%%%%%%%%%%%%%%%%%%%%%
%%%%%%%%%%%%%%%%%%%%%%%%%%%%%%%%%%%%%%%%%%%%%%%%%%%%%%%%%%%%
\begin{titlepage}

\vskip.5cm
\begin{center}

{\huge \bf  5D UED:  Flat and Flavorless}
\end{center}

\begin{center}
{\bf  {Csaba Cs\'aki}$^a$, {Johannes Heinonen}$^a$, {Jay Hubisz}$^b$, \\
{Seong Chan Park}$^c$, and {Jing Shu}$^c$}\\
\end{center}
\vskip 8pt

\begin{center}
{\it  $^{a}$ {\it Institute for High Energy Phenomenology\\
Newman Laboratory of Elementary Particle Physics\\
Cornell University, Ithaca, NY 14853, USA } \\
\vspace{.1in}
$^b$ {\it Syracuse University, 201 Physics Building, Syracuse, NY  13244 USA}\\
\vspace{.1in}
$^c$ {\it Institute for the Physics and Mathematics of the Universe, The University of Tokyo, Chiba $277-8568$, Japan
} }

\vspace*{0.2cm}

{\it E-mail:} {\tt  csaki@cornell.edu, jh337@cornell.edu, jhubisz@physics.syr.edu, seongchan.park@ipmu.jp, jing.shu@ipmu.jp }
\end{center}

\vglue 0.3truecm

\begin{abstract}
\vskip 3pt \noindent
5D UED is not automatically minimally flavor violating.  This is due to flavor asymmetric counter-terms required on the branes.  Additionally, there are likely to be higher dimensional operators which directly contribute to flavor observables.   We document a mostly unsuccessful attempt at utilizing localization in a flat extra dimension to resolve these flavor constraints while maintaining KK-parity as a good quantum number.  It is unsuccessful insofar as we seem to be forced to add brane operators in such a way as to precisely mimic the effects of a double throat warped extra dimension.   In the course of our efforts, we encounter and present solutions to a problem common to many extra dimensional models in which fields are ``doubly localized:" ultra-light modes.   Under scrutiny, this issue seems tied to an intrinsic tension between maintaining Kaluza-Klein parity and resolving mass hierarchies via localization.
\end{abstract}
\end{titlepage}

\newpage

%%%%%%%%%%%%%%%%%%%%%%%%%%%%%%%%%%%%%%%%%%%%%%%%%%%%%%%%%%%%
%%%%%%%%%%%%%%%%%%%%%%%%%%%%%%%%%%%%%%%%%%%%%%%%%%%%%%%%%%%%
\section{Introduction}
\label{intro}
\setcounter{equation}{0}
\setcounter{footnote}{0}

Extra dimensions have found many uses over the past decade. One interesting possibility was to use fermion wave function overlaps along the extra dimension to explain the fermion masses and mixing angle hierarchies. While such constructions have originally been proposed for large extra dimensional models~\cite{ArkaniHamed:1999dc}, they have turned out to be  most useful in the context of warped extra dimensions~\cite{Grossman:1999ra,Gherghetta:2000qt,Huber:2003tu}. Another important development was the realization that there could be a geometric origin for the stability of a weakly interacting massive particle (WIMP) dark matter via Kaluza-Klein (KK) parity \cite{UEDDM1, UEDDM2}. This $Z_2$ symmetry is implemented as a reflection around the midpoint of the extra dimension and is a remnant of 5D translational invariance, which is broken by the appearance of the end points of the interval (equivalently the orbifold fixed points in the orbifold language). 

KK parity has been the basic ingredient of the construction called Universal Extra Dimensions (UED), where all the Standard Model (SM) fields live in a flat extra dimension, and all SM wave functions are  flat~\cite{Appelquist:2000nn}. The main consequence of UED is that due to the unbroken KK parity the lightest KK odd particle (LKP) is a stable DM candidate~\cite{UEDDM1,UEDDM2} (usually it is the lightest KK mode of the $U(1)_Y$ gauge boson, $B_1$), and electroweak precision constraints are greatly reduced, the KK scale can be even below the TeV scale~\cite{Appelquist:2000nn}. UED is usually also considered to be an example of minimal flavor violation (MFV), since the only source of flavor violation are the Yukawa couplings~\cite{Buras:2002ej,Buras:2003mk}. 
We should however emphasize that  many of the nice properties of UED (sub-TeV EW precision bounds, MFV) are not a consequence of KK-parity, but due to complete KK-number conservation at tree level, which is due to the assumption of completely flat wave function and no localized terms.

While flat wave function profiles seem to be essential for the sub-TeV EWP bounds and MFV, it is only KK-parity that is necessary for the DM candidate. One may then wonder whether a non-trivial flavor structure would be possible along the lines of split fermions~\cite{ArkaniHamed:1999dc,Kaplan:2001ga}, while maintaining KK-parity. The main question we would like to address in this paper is whether a successful flavor model could possibly be based on a flat extra dimension with KK-parity, in analogy with those constructed in a warped geometry~\cite{Csaki:2008zd,otherRSflavor1,otherRSflavor2,otherRSflavor3,otherRSflavor4,otherRSflavor5,otherRSflavor6,otherRSflavor7}

In this case a KK-parity odd mass term for the fermions has to be added \cite{Park:2009cs,suedDM1, suedDM2,suedCollider, sued4gen}, which can be thought of as a consequence of a domain wall inside the extra dimension \cite{light1}. If we then want hierarchical fermion wave functions, there are two possibilities: the fermions could either be peaked at the center or at the boundaries. However, we find that there are always light modes trapped at the domain wall in the center, so only the second possibility can be realized. This will yield another problem: if the fermions are in the center, the Higgs needs to be close to the boundaries to get the flavor hierarchies from the wave function overlaps.  There are extra spontaneously broken approximate global symmetries in this ``doubly-localized" scenario, leading to extra KK-parity odd pseudo-Goldstone bosons.  We explain how to get around this problem through the 5D top-quark contribution to the Coleman-Weinberg potential~\cite{Coleman:1973jx}. 
%We expect that a similar construction is necessary in models with two-throat warped spaces as well.

The paper is organized as follows. In Section~\ref{sec:fermions} we present the setup with a flat extra dimension with KK-parity and exponential non-trivial fermion wave funcions, and explain in detail the origin of the ultra-light fermions.  In Section~\ref{sec:naivehiggssetup} we explain the origin and the properties of pseudo-Goldstone bosons with a boundary localized Higgs, and show how loop effects lift these Goldstone bosons.  In Section~\ref{sec:flavor} we analyze the flavor constraints of this model and show that the realistic setups are basically low-energy effective theory versions of a two-throat warped model. We explain in detail how to establish the proper correspondence. Finally, we analyze indirect constraints on the model in Section~\ref{sec:EWPC} and conclude in Section~\ref{sec:conclusion}.
Details of the calculation of the fermion spectrum are given in Appendix~\ref{app:fermions}. In Appendix~\ref{sec:newsetup} we show how to construct an alternative model with a flat Higgs wave avoiding tension with electroweak precision bounds.

%%%%%%%%%%%%%%%%%%%%%%%%%%%%%%%%%%%%%%%%%%%%%%%%%%%%%%%%%%%%
%%%%%%%%%%%%%%%%%%%%%%%%%%%%%%%%%%%%%%%%%%%%%%%%%%%%%%%%%%%%
\section{Localizing Fermions with KK-parity} \label{sec:fermions}
Let us consider a bulk 5D fermion with a varying bulk mass.  The geometry of the extra dimension is flat, and the extra dimensional coordinate, $y$, varies between two endpoints, $y=\pm L/2$:
\begin{equation}
        \label{eq:BulkAction}
 S = \int d^4 x \int_{-L/2}^{+L/2} dy
\left[\frac{i}{2}
\bar{\Psi}\, \Gamma^M \overleftrightarrow{\partial}_M \Psi
 - m_5(y) \bar{\Psi} \Psi  \right],
\end{equation}
where $\bar\Psi \overleftrightarrow{\partial_M} \Psi = \bar\Psi \partial_M \Psi - (\partial_M \bar\Psi) \Psi$ and the  gamma matrices are $\Gamma^M= (\gamma^\mu, i\gamma^5)$.
We introduce a $y$-dependent fermion mass which is a step-function that changes sign at the midpoint of the extra-dimension:
\begin{eqnarray}
m_5(y) = \mu ~\epsilon(y) = \left\{ \begin{array}{rl} -\mu, & y<0 \\Ê+\mu, & y >0Ê\end{array} \right.
\end{eqnarray}
Naively, this violates KK-parity, as the fermion mass changes sign across the midpoint of the extra dimension.  However, the \emph{intrisic} KK-parity of the 5D Dirac fermion compensates for the sign flip of the bulk mass under a KK-parity transformation.  Constant (or in general, even under KK-parity) fermion bulk masses which are sterile under KK-parity are forbidden for this reason~\cite{Park:2009cs}.

%%%%%%%%%%%%%%%%%%%%%%%%%%%%%%
\subsection{Zero mass and ultralight Modes}
The details of the equations of motion and the mass spectrum are given in Appendix~\ref{app:fermions}.
For $n=0$, there are massless solutions, or zero modes, with wave functions
\begin{eqnarray}
 f_0 (y) = \sqrt{\frac{+ \mu}{1-e^{- \mu L}}} e^{- \mu |y|} ,~~~~~~~g_0 (y) = \sqrt{\frac{- \mu}{1-e^{+ \mu L}}} e^{+ \mu |y|},
\end{eqnarray}
where $f_0$ denotes the right-handed (RH) and $g_0$ the left-handed (LH) solution. The wave functions of these zero modes are governed by the boundary conditions (which determine whether the zero mode is LH or RH) and the sign of $\mu$.  For example, if $\mu>0$, and the zero mode is RH, $f_0$ is peaked towards the mid-point of the extra dimension ($y=0$).  There are four possible cases, which we display in Table~\ref{tab:zeromodes}.
\begin{table}
\center{\begin{tabular}{|l|l|l|l|}
\hline
$\mathrm{sign} (\mu)$ & Chirality & Localization & Ultralight KK-Mode \\
\hline
$\mu > 0$ & RH & Midpoint & No \\
$\mu < 0$ & RH & Endpoints & Yes \\
$\mu > 0$ & LH & Endpoints & Yes \\
$\mu < 0$ & LH & Midpoint & No \\
\hline
\end{tabular}}
\caption{\label{tab:zeromodes} In this table, we give the localization of fermions for the different possible orbifold boundary conditions that produce either RH or LH zero modes, and for the different signs of the varying fermion bulk mass profile}
\end{table}

As we will discuss in further detail below, the localization of the zero mode determines the mass of the first KK-mode.  We note in particular that in each case where there is a zero mode which is ``doubly-localized" (where the wave function is sharply peaked towards each of the end-points of the extra dimension), there is a KK-mode whose mass is highly suppressed compared with the inverse size of the extra dimension. 

%%%%%%%%%%%%%%%%%%%%%%%%%%%%%%
%\subsection{KK modes}

%

	The boundary conditions and bulk equations of motion allow for at most one non-zero mass eigenvalue with a mass smaller than $\mu$, determined by $m_1^2 = \mu^2- \kappa_1^2$, with
	\begin{equation} \begin{split}
		 \kappa_n = \mp \mu \tanh \frac{\kappa_n L}{2} \qquad \text{ (for RH/LH zero mode),} 
	\end{split} \end{equation}
	where the sign is set by the boundary conditions, allowing for either a RH or LH zero mode.
	This equation only has solutions for $\mp \mu > 0$. Under this assumption, and if $\mu L \gg 1$, the corresponding mass eigenvalue is then given by\footnote{For large $x$: $\tanh x \approx 1 - 2 e^{-2 x}$} $m_1^2 = \mu^2 - \kappa_1^2 \approx 4 \mu^2 e^{-\mu L}$. For $L \sim 1/\text{TeV} $ these light modes would be problematic and we choose boundary conditions to avoid them in our attempt to build a model of UED flavor.

\begin{figure}[t]%[htbp]
\centering
\includegraphics[width=0.85 \textwidth]{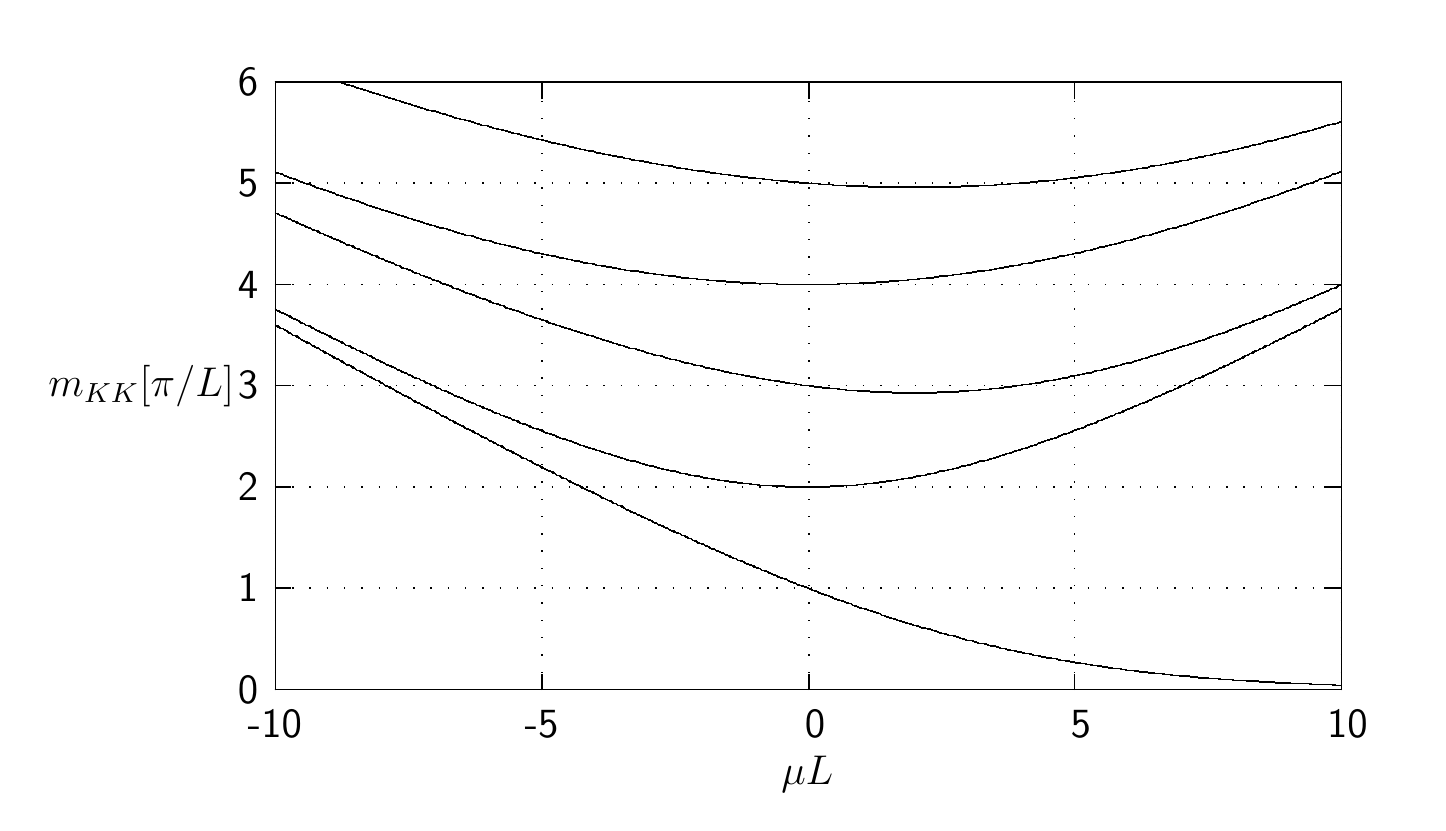}
\caption{\label{Fig:spectrum} First five KK-masses in the spectrum for a 5D fermion with a LH zero mode.  Note the existence of an ultralight mode in the case that $\mu$ is large (compared with $1/L$) and positive.}
\end{figure}

The complete mass spectrum is derived in Appendix~\ref{app:fermions}. The dependence of the spectrum on $\mu$ is shown in Figure~\ref{Fig:spectrum}, where a LH zero mode is assumed.  For negative $\mu$, the LH zero mode is localized towards the center of the extra dimension, at the kink in the bulk mass.  However, when $\mu$ is positive, this zero mode is localized towards the endpoints.  In this latter case, there is also an ``ultra-light" mode whose mass is exponentially suppressed, as derived above.
Some of these ultralight modes will be lifted by EWSB, as explained in Appendix~\ref{app:ultralightEWSB}. Nevertheless, there will always remain ultra-light fermions in the spectrum when the zero mode is localized in the center of the extra dimension.
%%%%%%%%%%%%%%%
%%%%%%%%%%%%%%

%%%%%%%%%%%%%%%%%%%%%%%%%%%%%%
%\subsection{Ultra-light Modes}

\pagebreak
\begin{center} {\bf Physical explanation of the ultra-light mode} \end{center} \label{sec:ultralight}

It is relatively easy to understand the presence or absence of the light first KK-mode.  It is essentially due to the usual domain wall trapping of chiral modes, a mechanism which functions regardless of the choice of boundary conditions in a finite extra dimension.

To elaborate, let us assume that we are looking at a bulk mass $m_5(y) = \mu\epsilon(y)$, where $\mu >0$. In the case of a doubly infinite extra dimension (the limit $L \rightarrow \infty$), this domain wall traps a right handed field.  This right-handed field is massless, since its corresponding left-handed partner is not normalizable.  If we now compactify this extra dimension, we must impose boundary conditions.  These boundary conditions can in principle  allow for either a left- \emph{or} right handed zero mode, that \emph{a priori} has nothing to do with the mode that is trapped at the domain wall.  In the discussion that follows, we track the fate of the trapped RH domain wall fermion when $L$ is taken to be finite.

First, let us assume that we impose BC that allow for a right-handed zero mode, i.e. $g_n |_{y=\pm L/2}=0$. In this case, the zero mode allowed by the boundary conditions is precisely the trapped mode of the inifinite extra dimension:  $f_{0} (y) \propto e^{-\mu |y|}$.  In this case, the spectrum contains no other low-lying modes.  

If we instead impose BC that permit only a left-handed zero mode, i.e. $f_n |_{y=\pm L/2} =0$, the true zero mode is $g_{0}(y) \propto e^{+ \mu |y|}$ (localized at the boundaries) and all other modes get a mass.  This mode would be non-normalizable in the case of an infinite extra dimension.  For sufficiently large $\mu \gg L$ the ``would-be zero mode" $f (y) \propto e^{-\mu |y|}$, localized on the domain-wall in the infinite $L$ limit, is exponentially close to satisfying the required boundary conditions that give the LH zero mode.  Hence the first KK-mode is made out of a right-handed ``trapped" mode (and an associated left-handed mode, which is localized towards the boundaries, but has a KK-odd wave-function).  The mass of this Dirac fermion is exponentially suppressed by $\sim e^{-\mu L/2}$.  In the large $L$ limit, both the LH zero mode and the LH component of the light KK mode become non-normalizable, and the RH trapped mode becomes the lowest lying physical excitation, the usual trapped chiral mode in large extra dimensions.

Some of these ultralight modes are actually lifted by the Higgs mechanism, as we explain in Appendix~\ref{app:ultralightEWSB}.  However, for each SM fermion, there is still a remaining light KK-odd mode, with mass lower than that of its SM partner.

%%%%%%%%%%%%%%%%%%%%%%%%%%%%%%%%%%%%%%%%%%%%%%%%%%%%%%%%%%%%
%%%%%%%%%%%%%%%%%%%%%%%%%%%%%%%%%%%%%%%%%%%%%%%%%%%%%%%%%%%%
\section{Boundary localized Higgs and ultra-light scalars} \label{sec:naivehiggssetup}

As we saw in the previous section, we require that the fermions be localized in the middle of the extra dimension to avoid light fermionic modes.  One way to get the hierarchical overlap with the Higgs required to naturally generate the hierarchy of fermion masses is to localize the Higgs in a KK-parity symmetric way on the endpoints of the interval.  In this section we show that such a scenario generates new ultralight KK-odd scalar fields,  which are lifted by a one-loop Coleman-Weinberg potential that is generated by 5D fermion and gauge boson loops.
%necessitating further model building to remove them from the spectrum. 

%%%%%%%%%%%%%%%%%%%%%%%%%%%%%%%%%%%%%%%%%%%%%%%%%%%%%%%%%%%%
\subsection{Higgs Model Setup}

We start with a complex scalar field $\Phi$ living in the bulk with a bulk potential and two identical  boundary potentials which respect KK parity.  This model has been studied previously in~\cite{Haba:2009uu}, and we summarize the results.  The scalar action is given by:
\begin{equation}
 S = \int d^4x\int_{-L/2}^{L/2}dy \left[ \frac{1}{L} |D_M\Phi|^2
     -\frac{1}{L} \mathcal{V}(\Phi)- \delta(y-L/2)V_{\frac{L}{2}}(\Phi)-\delta(y+L/2)V_{-\frac{L}{2}}(\Phi) \right].\label{p1}
\end{equation} 
We take the bulk potential to consist of a positive mass$^2$ term, while we assume boundary localized potentials which can lead to the development of a VEV (the two potentials are identical in order to respect KK-parity):
\begin{eqnarray}
 \mathcal{V}(\Phi) & = & m^2|\Phi|^2,  \label{bup} \\
 V_{-\frac{L}{2}}(\Phi) =   V_{\frac{L}{2}}(\Phi) & = &  \frac{\lambda}{4} \left( |\Phi|^2-\frac{v_0^2}{2} \right)^2.  \label{bp}
\end{eqnarray} 
The bulk equations of motion for the radial and Goldstone modes and the associated boundary conditions follow from the variation of this action with respect to the Higgs scalar field.  

%%%%%%%%%%%%%%%%%%%%%%%%%%%%%%
%\subsubsection{The Higgs VEV}
\begin{center} {\bf The Higgs VEV} \end{center}
The general solution for the VEV of the neutral component of $H$ is
 \begin{eqnarray}
 \langle H_0 \rangle = v(y)/\sqrt{2}=A\cosh(m y)+B \sinh(m y),
\end{eqnarray}
where the coefficients are determined by the boundary conditions.  There are multiple extrema, however the global minimum, or true vacuum, corresponds to a VEV profile which is KK-even.  The coefficients corresponding to this solution are 
\begin{equation}
A = \sqrt{\frac{ \lambda L v_0^2 c_h - 4 m s_h }{\lambda  L{c_h}^3}} \qquad \text{ and } \qquad B = 0,
\end{equation}
where ${c_h}=\cosh m L /2$ and ${s_h} = \sinh m L /2$.  We note that the KK-odd local minimum becomes degenerate with the KK-even vacuum in the limit as $m \rightarrow \infty$.  Obtaining the correct $W^\pm$ and $Z$ boson masses from this VEV fixes $v_0$ in terms of $m$ and $L$.

%%%%%%%%%%%%%%%%%%%%%%%%%%%%%%
\begin{center} {\bf Light scalar spectrum} \end{center}
Since the bulk equation of motion is linear, the 
%radial 
fluctuations about the VEV obey the same bulk equation of motion as the VEV, and therefore the general form of the solution is the same:
 \begin{eqnarray}
h^{(n)}_0(y)=A \cosh (k_n y)+ B \sinh (k_n y),
\end{eqnarray}
where $k_n=\sqrt{m^2-m_n^2}$ (for exponential localization, all light modes obey $m_n \ll m$).  Imposing the boundary conditions enforced by the brane-localized potentials, we find that the mass eigenvalues are given by
\begin{equation} \begin{split}
 k_0 \tanh (k_0 L/2)  &= 
 %\left( 
 3 m \tanh m L/2 - \frac{\lambda L v_0^2}{2}  
 %\right) 
 \qquad \text{ (KK-even SM Higgs)} \\
 \frac{k_1}{\tanh (k_1 L/2)}  &=   
 %\left( 
 3 m \tanh m L/2 - \frac{\lambda L v_0^2}{2}  
 %\right) 
  \qquad \text{ (KK-odd partner)}.
\end{split} \end{equation}

This will allow for a light KK-even Higgs mode if the bulk mass and the boundary mass are chosen appropriately.  We note that obtaining a weak-scale Higgs mechanism from a TeV$^{-1}$ size extra dimension requires tuning of the boundary potential negative mass$^2$ against the positive bulk mass$^2$.  There is also a light KK-odd solution which is degenerate with the KK-even Higgs in the limit $m L \gg 1$.

While the possibility of a weak-scale KK-odd Higgs field is quite interesting for phenomenological reasons, we find that there are also ultralight \emph{physical} KK-odd Goldstone fluctuations which are dangerous from the perspective of various constraints on such modes.  The spectrum of Goldstone fluctuations is given by:
\begin{equation} \begin{split}
m_0 & =   0			   \hspace{2.65cm} \text{(KK-even zero mode)} \\
\frac{k_1}{\tanh (k_1 L/2)} & =  m \tanh \frac{m L}{2}   \qquad \text{(ultra-light KK-odd mode)}
\end{split}  \end{equation}
For large $m \gg L^{-1}$, we can again expand the $\tanh$ to obtain an approximate formula $m_1^2 \sim 8 m^2 e^{-m L}$.  To obtain the observed fermion mass hierarchy, we would require $m L \sim 30$, pushing this mass down into the $1-10$~eV range.  In the next section, we provide a geometric origin for these light modes, and describe how quantum effects lift them above the weak-scale.

In the case of a Higgs electroweak doublet, the KK-even Goldstone bosons are the ones which become the longitudinal polarizations of the SM $W^\pm$ and $Z$ bosons, while the KK-odd modes are physical light scalar degrees of freedom in the low energy theory.  There are 3 such scalar fields in this model, $\Pi^\pm$ and $\Pi^0$.  The $Z$-boson would decay rather efficiently through the $Z\rightarrow \Pi^+ \Pi^-$ channel  should there be available phase space (there is no coupling of the $Z$ to two $\Pi^0$'s, as these fields all correspond to $U(1)$ generators).  The ultralight $\Pi^0$ is also quite dangerous.  It is in thermal equilibrium at weak-scale temperatures, but has no way in which it can annihilate efficiently, as it is less massive (at tree level) than any particle in the SM except the photon and neutrinos.  We soon show that such dangerous modes are lifted well above the weak scale, avoiding such problems.

We explain in the next section that the charged KK-odd pseudo-Goldstones are naturally lifted by the SM gauge and Yukawa interactions, which explicitly break the global symmetries protecting them.  While Yukawa interactions break all the extra global symmetries, gauge interaction preserve one global $U(1)$ symmetry, resulting in a splitting between the charged and neutral KK-odd Goldstone bosons.

%%%%%%%%%%%%%%%%%%%%%%%%%%%%%%%%%%%%%%%%%%%%%%%%%%%%%%%%%%%%
\subsection{ KK-odd (pseudo-)Goldstone bosons }
\label{sec:KKoddGoldstone}

\begin{center} {\bf A symmetry argument for their origin} \end{center}

The origin of the light scalar modes is easiest to understand in the language of a deconstructed extra dimension.  For our purposes, we require only a 2-site model.  On each of these two sites, we place Higgs fields $H_1$ and $H_2$.  We do not add any terms in the Lagrangian which contain both $H_1$ and $H_2$ (analogous with requiring 5D locality).   This is an appropriate simplified picture of the model in which the Higgs is strongly localized on the boundaries.  Before adding any gauge interactions, this setup has an extended global symmetry structure of one $SU(2) \times U(1)$ global symmetry on each site, which is broken spontaneously by the VEVs of each of this Higgs fields to $U(1)$\footnote{The actual global symmetry breaking pattern is $SO(4) \rightarrow SO(3) \sim SU(2)_\text{custodial}$, but it is sufficient, and perhaps more clear, to consider $SU(2) \times U(1)$, as this is the part that will be gauged, breaking the full global $SO(4)$ explicitly.  The Goldstone boson counting argument that follows is the same in either case.}.

When the two Higgs fields acquire VEVs, they break their own global symmetries down to one $U(1)$ each, resulting in six Goldstone bosons. These are most conveniently grouped into three KK-even and three KK-odd Goldstone bosons (symmetric and anti-symmetric combinations, respectively, of the Goldstone bosons from each site).  

\begin{figure}[t]
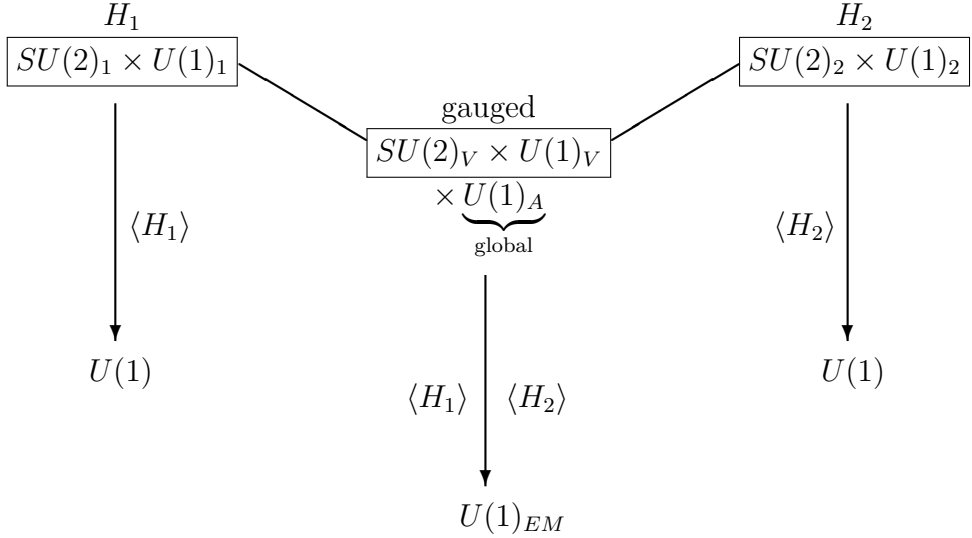
%[htbp]
\makebox[0.75 \textwidth][t]{}
\put(-170,130){$\begin{array}{c} %Q_L \\ t_R \\ 
\text{gauged} \\  \fcolorbox{black}{white}{$SU(2)_V \times U(1)_V$} \\ \times \underbrace{U(1)_{A}}_{\text{global}} 
\end{array}$ }
\put(-310,180){ $\begin{array}{c} H_1 \\ \fcolorbox{black}{white}{$SU(2)_1 \times U(1)_1$} \end{array}$ }
\put(-33,180){ $\begin{array}{c} H_2 \\ \fcolorbox{black}{white}{$SU(2)_2 \times U(1)_2$} \end{array}$ }
\put(-255,110){$\left< H_1 \right>$}
\put(-150,45){$\left< H_1 \right> \quad \left< H_2 \right>$}
\put(-11,110){$\left< H_2 \right>$}
\put(-270,55){$U(1)$}
\put(-130,0){$U(1)_{EM}$}
\put(7,55){$U(1)$}
\thicklines
\put(-217,175){ \line(5,-3){48} }
\put(-28,175){ \line(-5,-3){48} }
\put(-260,160){\vector(0,-1){90}}
\put(-120,95){\vector(0,-1){80}}
\put(17,160){\vector(0,-1){90}}
\caption{\label{fig:deconstruct} Deconstructed  analog of the 5D scalar field model in Sec.~\ref{sec:naivehiggssetup}.}
\end{figure}

In usual deconstruction, gauge groups reside at each site, with sigma fields linking them.  These sigma fields acquire VEVs, spontaneously breaking each pair of sites to the diagonal subgroup.  In our case, it is sufficient to imagine the VEV of this sigma field being taken to infinity, decoupling the antisymmetric combination of the gauge fields completely.   The full picture of our toy model is then to gauge only the diagonal subgroup of the global symmetries on the two sites, furnishing the SM gauge group.  We illustrate the complete setup in Figure~\ref{fig:deconstruct}.
 Now when the Higgs fields acquire VEVs, the gauge group is broken to $U(1)_\text{em}$, and 3 of the six Goldstone bosons must be eaten, leaving 3 physical massless scalar fields.  Adding exponentially suppressed mass mixing terms between $H_1$ and $H_2$ mimics the exponential suppression of the wave function in the middle of the extra dimension, and explicitly breaks the anti-symmetric combination of the global symmetry generators.  This leads to exponentially small masses for the (now pseudo-)Goldstone bosons corresponding to those generators.  These pseudo-Goldstone bosons are to be identified with the ultralight fields of the previous section.

\pagebreak
\begin{center} {\bf Explicit breaking - masses through quantum effects} \end{center}

Gauging the diagonal subgroup of the $\left[ SU(2) \times U(1)\right]^2$ global symmetry group breaks some of these global symmetries explicitly.  Specifically, the antisymmetric combination of the two sets of $SU(2)$ generators do not form a full $SU(2)$ group.   The original $SU(2) \times SU(2)$ global symmetry is thus broken to diagonal vector subgroup, the gauged SM $SU(2)_L^V$.  Due to the triviality of $U(1)$ symmetries, the antisymmetric combination of $U(1)$'s is not broken by gauging the diagonal subgroup.  Taking into account the explict breaking, the electroweak symmetry breaking pattern is now
\begin{equation}
\left[ SU(2)_L^V \times U(1)_Y^V \right] \times U(1)^A \rightarrow \left[ U(1)_\text{em} \right],
\end{equation}
where the gauged groups are placed in brackets.  We see that there will be 3 eaten Goldstone bosons corresponding to the gauged generators, and one remaining neutral KK-odd physical Goldstone from the breaking of the KK-odd global $U(1)^A$.    The two charged KK-odd Goldstone bosons have been lifted by gauging the diagonal subgroup of the full set of global symmetries.  Explicitly, loops involving $SU(2)_L$ gauge bosons will generate masses for the charged pseudo-Goldstone bosons.  In the unbroken phase, the symmetry breaking operator is quartic in the Higgs doublets, and the diagram is logarithmically divergent in the effective theory (and cutoff by the KK-scale in the full 5D theory).

Fortunately, Yukawa couplings further reduce the symmetry, breaking the global $U(1)^A$ as well.  If one includes, in the low energy effective theory, a $Q_L$ doublet and $t_R$ singlet with gauge invariant and KK-parity symmetric Yukawa couplings with the two Higgs fields, the complete set of anti-symmetric generators is broken.  This is easy to see from the form of the Yukawas in our low energy construction:
\begin{equation}
\mathcal {L}_\text{top} = \frac{1}{\sqrt{2}} \lambda_t \bar{t}_R (H_1 + H_2)^T i \tau^2 Q_L + \text{h.c.}
\end{equation}
Such an interaction can only be invariant under transformations that are identical for $H_1$ and $H_2$ (the diagonal subgroup of the full $\left[ SU(2) \times U(1) \right]^2$).  In the unbroken phase, quantum effects involving the top-quark generate the following terms in the one-loop effective potential:
\begin{equation}
\mathcal{V}_\text{1-loop} = + \frac{3 \lambda_t^2}{32 \pi^2} \left[ \left(H_1^\dagger H_1+ H_2^\dagger H_2 \right) \Lambda^2  +\left(H_1^\dagger H_2 + H_2^\dagger H_1\right) \Lambda_{1,2}^2 \right].
\end{equation}
We distinguish the cut-offs for the two different types of terms since they have different meanings in the 5D picture that UV completes our simple model.  The first terms are local in 5D, in that they involve the Higgs field only on one brane.  These quantum effects are the precursor to renormalization of the 5D mass term.  The second set of terms, however, are non-local.  In the full extra-dimension, such loops can not be divergent, and are expected to be cut off at the inverse size of the extra dimension.
When we expand the operators that mix $H_1$ and $H_2$, we find mass terms for the KK-odd Goldstone bosons.  

\begin{figure}[t]%[htbp]
\centering
\includegraphics[width=0.65 \textwidth]{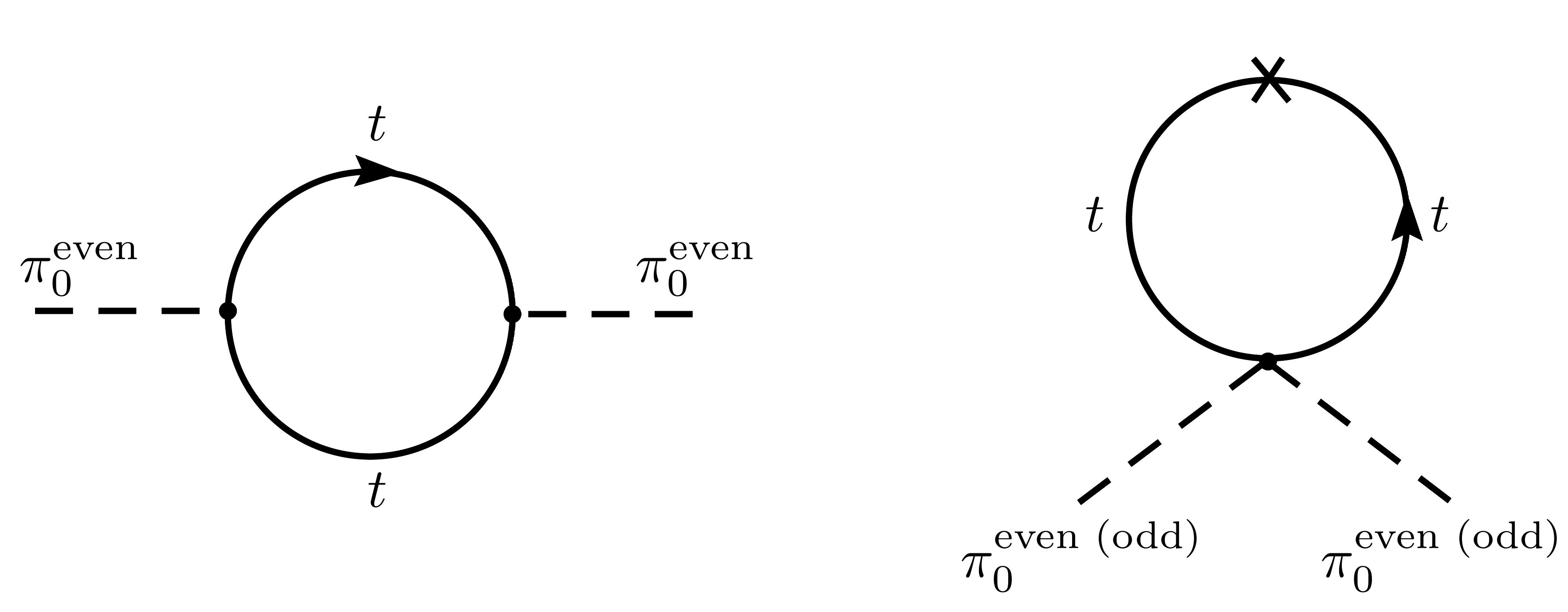}
\caption{\label{fig:goldstonediagrams} Diagrams contributing to the two point function for the KK-(odd) even (pseudo)-Goldstone bosons.  For the KK-even Goldstone, which is eaten by the $Z$-boson, the two diagrams precisely cancel, while for the KK-odd pseudo-Goldstone boson, only the second diagram contributes, giving a positive contribution to the mass$^2$.}
\end{figure}

These mass$^2$ terms are positive, as there is a sign arising in the expansion of the Higgs field about its vev that cancels the minus sign due to fermion statistics.   In the broken phase, the origin of the positive mass$^2$ for the KK-odd pseudo-Goldstone boson is from a quadratically divergent diagram, and the sign is easy to see from a simple symmetry argument.  In calculating the two point function of an exact Goldstone boson, there is a cancelation that occurs between two separate diagrams, as shown in Figure~\ref{fig:goldstonediagrams}.  The first diagram is negative, while the second is positive, to manifest the cancelation.  For the KK-odd pseudo-Goldstone, only the second of these two diagrams contributes, giving a positive uncanceled contribution to its mass.

Calculating the full 5D effect, assuming the light Higgs is completely localized on the branes and that the top-quark has a flat wave-function, we find masses for the KK-odd pseudo-Goldstone bosons:
\begin{equation}
m^2_\text{odd} = \frac{3 \lambda_t^2}{16 \pi^2} \frac{6 \zeta(3)}{L^2} \approx (400 \text{ GeV})^2 \times \left(\frac{1 \text{ TeV}^{-1}}{L}\right)^2
\end{equation}
well in line with the expectation that $\Lambda_{1,2} \sim M_\text{KK}$.
 
%%%%%%%%%%%%%%%%%%%%%%%%%%%%%%%%%%%%%%%%%%%%%%%%%%%%%%%%%%%%
%%%%%%%%%%%%%%%%%%%%%%%%%%%%%%%%%%%%%%%%%%%%%%%%%%%%%%%%%%%%
\section{Flavor constraints} \label{sec:flavor}

Now that we have developed a UED model which accomplishes the zeroth order goal of producing the hierarchy of SM fermion masses in a natural way, we set out to study the bounds imposed on this model by flavor physics.  Specifically, it has been shown  that warped geometry provides an automatic built-in GIM mechanism which suppresses tree-level flavor changing neutral currents~\cite{Csaki:2008zd,otherRSflavor1,otherRSflavor2,otherRSflavor3,otherRSflavor4,otherRSflavor5,otherRSflavor6,otherRSflavor7}.  Since this mechanism is due to the geometric warping in RS models, it should come as little surprise that flavor physics imposes rather strong constraints on this flat-space model, as we show in this section.

%%%%%%%%%%%%%%%%%%%%%%%%%%%%%%
\subsection{Flavor hierarchy and the CKM matrix}

In our model, the light fermions are constructed from LH and RH zero modes which are localized at the center of the extra dimension, on the domain wall.  In this arrangement, there are no ultra-light states since the zero modes are not doubly localized.  We arrange the 5D bulk fermions in terms of gauge eigenstates which carry flavor quantum numbers:
\begin{equation}
\Psi_Q = ( Q_1, Q_2, Q_3 ), \qquad \Psi_{u} = (u_1, u_2, u_3) \quad \text{ and} \quad \Psi_{d} = (d_1, d_2, d_3).
\end{equation}
Boundary conditions for these 5D Dirac fermions are chosen such that the $Q_i$ contain LH zero modes, and the $u_i$ and $d_i$ contain RH zero modes. 
Rewriting the bulk masses in units of the inverse size of the extra dimension, the zero mode wave functions are given by
\begin{eqnarray}
&&f^{(0)}(y) = \sqrt{\frac{1}{L}} \sqrt{\frac{ c_{R}}{1-e^{- c_{R}}}} e^{- c_{R} \frac{|y|}{L} } = \sqrt{\frac{1}{L}} e^{- c_{R} (\frac{|y|}{L}-\frac{1}{2})} f(c_{R}) \nonumber \\
&&g^{(0)}(y) = \sqrt{\frac{1}{L}} \sqrt{\frac{-c_{L}}{1-e^{ c_{L}}}} e^{ c_{L} \frac{|y|}{L} } = \sqrt{\frac{1}{L}} e^{ c_{L} (\frac{|y|}{L}-\frac{1}{2})} f(-c_{L})
\end{eqnarray} 
where $c_L <0$ and $c_R >0$.  We have also introduced a ``flavor function" $f(c)$, given by
\begin{equation}
f(c) = \sqrt{\frac{c}{e^{c}-1}}.
\end{equation}

Note that this flavor function is identical to the one obtained in RS flavor models (see e.g.~\cite{Csaki:2008zd}):
\begin{equation}
f_{RS}(c) = \sqrt{\frac{1-2 c_{RS}}{1-e^{(2c_{RS}-1) k \pi r_c}}} = \sqrt{\frac{1}{k \pi r_c}} \sqrt{\frac{\tilde c}{e^{\tilde c}-1}} \sim \frac{1}{\sqrt{30}} f(\tilde c),
\end{equation}
where $r_c$ is the RS radius, $k$ is the curvature of the warped extra dimension, and we have defined $\tilde c \equiv (2 c - 1) k \pi r_c$.

For the purposes of our flavor study, we assume that the Higgs VEV is completely localized, in the form of delta-functions on the boundaries.  This is reasonable, given the amount of localization necessary to achieve the fermion mass hierarchy.
\begin{equation} 
{\mathcal L}_Y  = \frac{1}{2} \int_{-L/2}^{+L/2}  \left( \langle H \rangle \bar \Psi_q Y_u \Psi_{u} + \langle H^* \rangle \bar \Psi_q Y_d \Psi_{d}  + h.c. \right)\left[ \delta(y+L/2) + \delta(y-L/2) \right] 
\end{equation} 
Inserting the zero mode profiles for the fermions yields the SM fermion mass matrices in terms of the brane-localized Yukawa couplings and the flavor functions:
\begin{equation} \begin{split}
M_u & = \frac{v}{\sqrt{2}} f_{q} Y_u f_{u^c} \\
M_d & = \frac{v}{\sqrt{2}} f_{q} Y_d f_{d^c},
\end{split} \end{equation}
where we used the shorthand $f_q = \mathrm{diag} [ f(-c_{q_1}),f(-c_{q_3}),f(-c_{q_3})]$ and similar for $f_{u}$ and $f_{d}$.
We now diagonalize the up and down mass matrices, as usual, by unitary rotations $M_{u,d} = U_{L\,u,d} m_{u,d}^{SM} U_{R\,u,d}^\dagger$, where $m_{u,d}^{SM}$ are diagonal.  Rotating the fermion fields into this basis yields the CKM matrix $V_{CKM} = U_{L\,u}^\dagger U_{L\,d}$. Since we take the $f_q$'s to be hierarchical, one can check by explicit calculation that the eigenvalues are of the form
\begin{equation} 
(m_{u,d}^{SM})_{ii} \sim \frac{v}{\sqrt{2}} (Y_{u,d})_{ii} f_{q_i} f_{u_i, d_i},
\end{equation}
with diagonalization matrices having elements of the order~\cite{Huber:2003tu}
\begin{equation}
|(U_L)_{ij}| \sim \frac{f_{q_i}}{f_{q_j}} \qquad \text{Êand} \qquad  |(U_R)_{ij}| \sim \frac{f_{u_i,d_i}}{f_{u_j,d_j}}, \qquad \text{ for } iÊ\leq j.
\end{equation} 
This implies that $|(V_{CKM})_{ij}| \sim{f_{q_i}}/{f_{q_j}}$.  Inputting the hierarchical structure of the CKM matrix then fixes the spectrum of the left-handed $f_{q_i}$'s as
\bea
\frac{f_{q_2}}{f_{q_3}} \sim \lambda^2, ~~~  \frac{f_{q_1}}{f_{q_3}} \sim \lambda^3 \ ,
\eea
where $\lambda \sim \sin{\theta_c} \sim 0.2$.
The fermion mass hierarchy then determines the right handed wave functions as evaluated on the branes at $y=\pm L/2$. The corresponding $f_{u_i,d_i}$'s are fixed to be~\cite{Amsler:2008zzb}
\begin{align}
\frac{f_{u_1}}{f_{u_3}} & \sim  \frac{m_u}{m_t} \frac{1}{\lambda^3}  = 6.88 \times 10^{-4} & 
\frac{f_{u_2}}{f_{u_3}} & \sim  \frac{m_c}{m_t}\frac{1}{\lambda^2} =  1.02 \times 10^{-1} & 
   &      \\
\frac{f_{d_1}}{f_{u_3}} & \sim  \frac{m_d}{m_t}\frac{1}{\lambda^3} =  1.84 \times 10^{-3} & 
\frac{f_{d_2}}{f_{u_3}} & \sim  \frac{m_s}{m_t} \frac{1}{\lambda^2} = 8.63 \times 10^{-3} & 
\frac{f_{d_3}}{f_{u_3}} & \sim  \frac{m_b}{m_t} = 1.76 \times 10^{-2}. \nonumber 
\end{align}
The strategy in the lepton sector is similar, although we do not attempt here to generate the PMNS structure.  Ignoring the mechanism of neutrino mass generation,  and assuming the left-handed charged leptons are flat  (providing the best-case-scenario for electroweak precision fits), we then obtain the flavor functions for the right-handed lepton sector:
\bea
f_{l_3^c} \sim \frac{m_\tau}{m_t} = 1.3 \times 10^{-2} ~~~ f_{l_2^c} \sim \frac{m_\mu}{m_t} =  7.75 \times 10^{-4}~~~ f_{l_1^c} \sim \frac{m_e}{m_t} =  3.74 \times 10^{-6}.
\eea

%%%%%%%%%%%%%%%%%%%%%%%%%%%%%%
\subsection{Contributions to FCNC's}

Generating the fermion mass hierarchy comes with a steep price.  As described above, the spectrum of fermion masses arises from fermion wave function overlaps with the brane localized Yukawa couplings.  However, the necessary variation in the fermion wave functions across generations leads to non-universality in the couplings of gauge-boson KK-modes to SM fermions in the gauge basis.  The diagonalization of these fermions to the mass basis then generates off-diagonal (in flavor-space) couplings of the KK-even gauge KK-modes with SM fermions.  The constraints on such couplings are very stringent, and put severe limits on models of UED flavor.

As the KK-gluons couple most strongly to light fermions, we focus on the flavor violating couplings of these fields.  There are similar, albeit somewhat weaker, constraints from exchange of $SU(2)_L$ and $U(1)_Y$ gauge boson KK-modes.  Due to KK-parity, only the KK-even modes contribute at tree-level.  Their wave functions, for $n>0$, are $G^{(2 n)}(y)=\sqrt{\tfrac{2}{L}} \cos \frac{2 \pi n y}{L}$.   In the original gauge eigenstate basis, before diagonalization to the mass basis, the couplings of the $2n^{th}$ KK-mode to a fermion zero mode with bulk mass $c/L$ are
\begin{equation} \begin{split} \label{eq:gcoupling}
g_s^{(2 n)}(c) 
		& = g^{4D} \sqrt{2}  
		\left[1- f_c^2 \, \gamma^{(2n)}_c\right], 
\end{split} \end{equation}
where the function $\gamma^{(2n)}_c$ is defined to be
\begin{equation}
\gamma^{(2n)}_c =  \frac{(-1)^n-1 + \left( \frac{\pi}{c} \right)^2 ( e^c-1) }{c (1 +   \left( \frac{\pi}{c} \right)^2)}.
\end{equation}
Note that this function is 1 for $c=0$ and  quickly becomes big and approaches $\frac{\pi^2 e^c}{c^3}$ already for small $c$ of order 5-10. We will see that it is this behavior, $\gamma^{(2n)}$ not of order 1, that will lead to unsuppressed flavor violation.

Since the $g (c)$'s are non-degenerate, we will find flavor off-diagonal couplings once we rotate to the mass basis
\begin{equation}
g^{(2n)}_{L\, u,d} \to U_{L\,u, d}^\dagger g_q^{(2n)} U_{L\,u,d} \qquad \text{and}Ê\qquad g^{(2n)}_{R\, u,d} \to U_{R\,u, d}^\dagger g_{u,d}^{(2n)} U_{R\,u,d}.
\end{equation}
The off-diagonal elements will be of order
%\begin{align}  
\begin{equation} \label{eq:offdiagcoupling}
(g_{L\, u,d})_{ij}  \sim  \sqrt{2} g_s \sum_k f_k^2 \, \gamma_k \left( \frac{f_{q_i}}{f_{q_k}} \right)^{\lambda_{ik}} \left( \frac{f_{q_j}}{f_{q_k}} \right)^{\lambda_{jk}}   ~~\text{with}~~\lambda_{n k} = \left\{ \begin{array}{l} +1~~~n < k \\ -1~~~n > k \end{array} \right.,
\end{equation}
where we have again assumed a hierarchy in the $f$ functions.  There are similar formulae for $(g_{R\, u})_{ij}$, and $(g_{R\, d})_{ij}$.

To compare these off diagonal couplings to constraints on generic operators which contribute to flavor changing neutral currents, we write down the effective Hamiltonian which arises at low energies after integrating out the KK-gluons.  We then calculate the coefficients of these operators and compare to experimental results.  Integrating out the KK gluon and performing some Fierz rearrangements of the operators yields the effective Hamiltonian for four fermion interactions~\cite{Csaki:2008zd}
\begin{eqnarray}
\mathcal{H} &=&\frac{1}{M_G^2} \left[
\frac{1}{6} g_L^{ij} g_L^{kl}
(\bar{q}_L^{i \alpha} \gamma_\mu q_{L \alpha}^j)\ (  \bar{q}_L^{k \beta} \gamma^\mu q_{L \beta}^l)
- g_R^{ij} g_L^{kl} \left( (\bar{q}_R^{i \alpha} q_{L
\alpha}^k)\
(\bar{q}_L^{l \beta} q_{R \beta}^j)
-\frac{1}{3}  (\bar{q}_R^{i \alpha} q_{L
\beta}^l)\
(\bar{q}_L^{k \beta} q_{R \alpha}^j)\right)\right]\nonumber\\  &=& C^1(M_G) (\bar{q}_L^{i \alpha} \gamma_\mu q_{L \alpha}^j)\ (  \bar{q}_L^{k \beta} \gamma^\mu q_{L \beta}^l) +  C^4(M_G)    (\bar{q}_R^{i \alpha} q_{L
   \alpha}^k)\
   (\bar{q}_L^{l \beta} q_{R \beta}^j) +   C^5(M_G) (\bar{q}_R^{i \alpha} q_{L
   \beta}^l)\
   (\bar{q}_L^{k \beta} q_{R \alpha}^j),
\nonumber
\end{eqnarray}
where $\alpha, \beta$ are color indices.

The coefficent $C^{4}_K (M_G)$, as measured in the Kaon system, has the strongest bound on contributions from new physics.   We thus use this variable as a rough check to see whether or not the UED flavor model is immediately ruled out.  In terms of our flavor functions, $f$, and couplings, $g$, the operator coefficient $C^4_K (M_G)$ arising from the exchange of a single level 2-n KK-gluon with mass $M_G^{2n}$ is given by:
\begin{equation} \begin{split}
C^4_{K}  & \sim \frac{2 g_s^2}{(M^{(2n)}_G)^2}  \left[ \Delta g_{q_1,q_2} ^{(2n)} \cdot \Delta g_{d^c_1,d^c_2} ^{(2n)} \right] \frac{ f_{q_1}}{f_{q_2}} \frac{ f_{d^c_1}}{f_{d^c_2}} \\
	& \sim  \frac{2 g_s^2}{(M^{(2n)}_G)^2}  \left[ \Delta g_{q_1,q_2} ^{(2)} \cdot \Delta g_{d^c_1,d^c_2} ^{(2n)} \right]  \frac{m_d}{m_s} \\
	& \equiv \left( \frac{ g_s}{M^{(2n)}_G} \xi \right)^2 
\end{split} \label{eq:C4} \end{equation}
where $\Delta g_{i,j}^{(2n)} \equiv g_{i}^{(2n)} -g_{j}^{(2n)}$.

If we plug in the values of the $c$-parameters for the left-handed and right-handed fermions required to match both the CKM structure and the fermion mass hierarchy, we find that $\xi \sim 94$, and
\begin{equation}
C^4_K \sim \left[ \frac{1}{1000~\text{TeV}} \left( L \cdot 500~\text{GeV} \right) \right]^2
\end{equation}
The limits on the new-physics flavor scale are  $\text{Re}~C_K^4 \lsim (10^4~\text{TeV})^{-2}$ and $\text{Im}~C_K^4 \lsim (10^5~\text{TeV})^{-2}$.
These bounds, assuming there is no relative suppression of the imaginary part of this operator, imply that $L^{-1} \gsim 500~\text{TeV}$.  This is the main challenge for a successful model of flavor in TeV-scale UED.

%%%%%%%%%%%%%%%%%%%%%%%%%%%%%%%%%%%%%%%%%%%%%%%%%%%%%%%%%%%%
\subsection{Center ``Brane'' localized kinetic terms}
In trying to bring the flat space model into accord with flavor bounds, we first consult the relatively successful models of flavor that can be built in the RS geometry.  We then import the key features of the RS geometry flavor solution into our flat space model.

\begin{figure}[t]%[htbp]
\centering
\includegraphics[width=0.8 \textwidth]{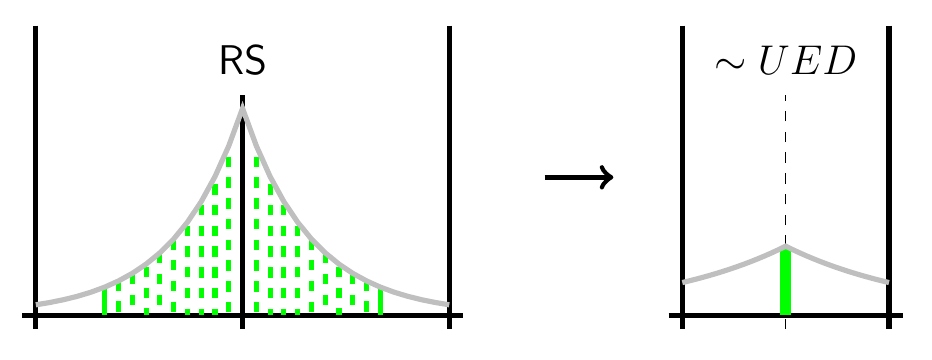}
\caption{\label{fig:doublethroat} Schematic picture how integrating out parts of a RS space yields  an effective ``UED'' space with brane localized operators at the center.}
\end{figure}

It is possible to reduce a warped space geometry to an approximately flat extra dimension by integrating out a large slice of the warped extra dimension.  The remaining warping is minimal and it is clear that this ``almost-flat-space" model will describe exactly the same physics as the complete warped extra dimension, encapsulating the RS-GIM mechanism in terms of brane localized operators which contain all effects of the modes which have been integrated out.  KK-parity can be realized in this scenario by beginning with a double warped throat~\cite{Cacciapaglia:2005pa,Agashe:2007jb} with an imposed discrete symmetry which interchanges the two throats.  We then integrate out the interior portion of this geometry, as illustrated in Figure~\ref{fig:doublethroat}.  Using this model as a guide, we find the set of operators that must be added in the flat-space model in order to suppress dangerous contributions to low-energy flavor observables.

%%%%%%%%%%%%%%%%%%%%%%%%%%%%%%
\subsubsection{Brane-localized operators in 5D UED}

We attempt to rectify the flat space UED model of flavor by adding kinetic terms for the 5D fermions on the domain wall in the center of the extra dimension.  It is intuitively clear that the addition of kinetic terms will alleviate the flavor bounds.  The addition of large fermion kinetic terms localized on the discontinuity will cause the zero-mode fermion couplings to gauge boson KK-modes to be more universal, suppressing the off-diagonal couplings in the fermion mass basis. 

The operators we add are of the form:
\begin{equation}
S_\text{fermion} = \int d^5x \left\{ \frac{i}{2}
 \bar{\Psi}\, \Gamma^\mu \overleftrightarrow{\partial}_\mu \Psi  \right\} \kappa_f L \delta(y).
\end{equation}
As we are only interested in zero mode fermions the derivation of the wave functions is especially easy as $m_0 = 0$. The strength of the BLKO, $\kappa_f$, appears only in the normalization of the fermion wave functions, and in the expression for the effective gauge coupling.   The flavor functions $f(c)$ in the presence of a brane kinetic term  are replaced by 
\begin{equation}
f(c,\kappa) = \sqrt{\frac{c}{(1+ c \kappa)e^c-1}}.
\end{equation}
%%
%\end{itemize}

%\noindent

\begin{figure}[t]%[htbp]
\centering
\includegraphics[width=0.4 \textwidth]{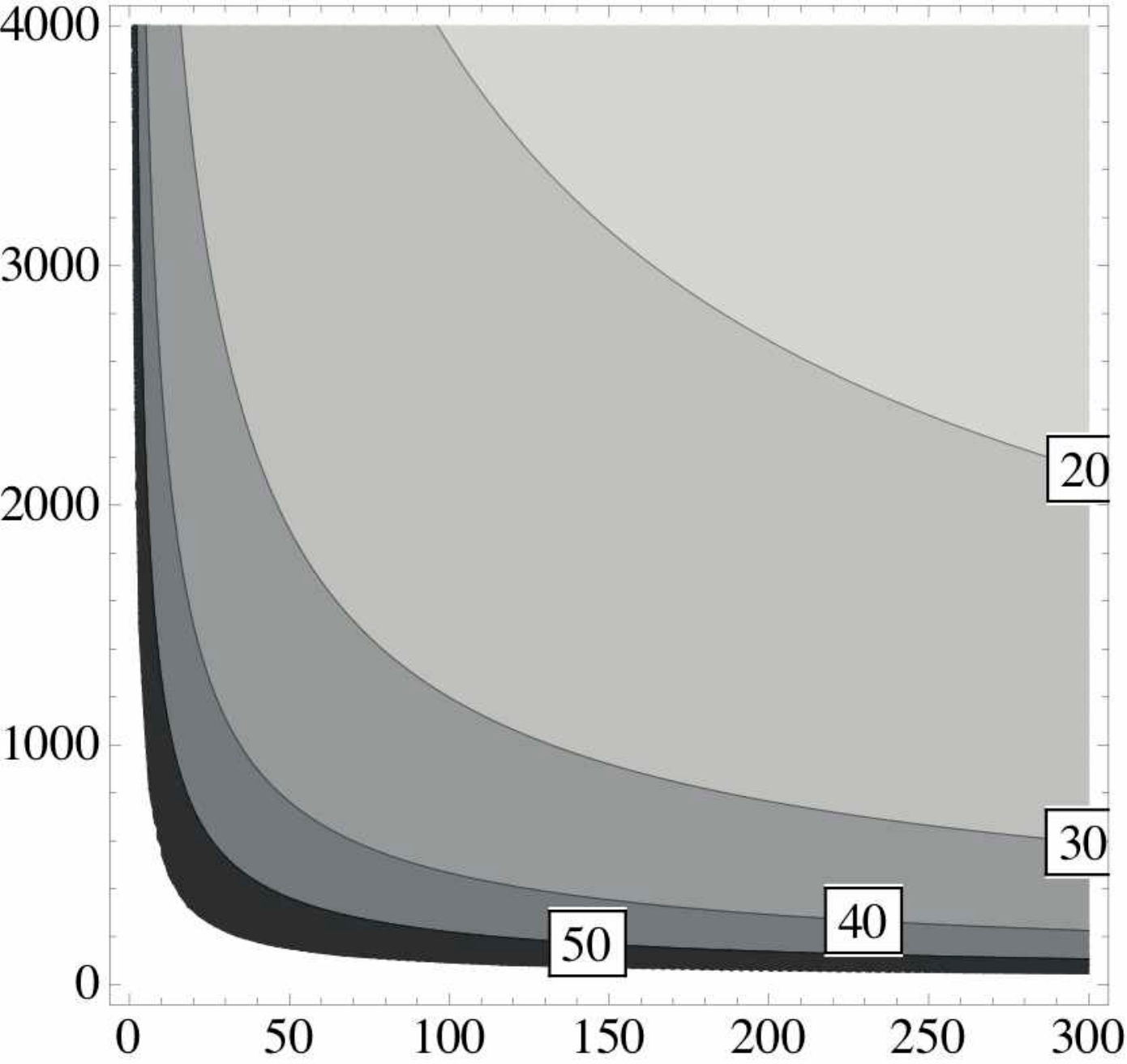}
\put(-97,155){\fcolorbox{black}{white}{$\kappa_{q_1}=\kappa_{d_1}=500$} }
\put(-90,-10){$\kappa_{q_2}$}
\put(-200,100){$\kappa_{d_2}$}
\hspace{1cm}
\includegraphics[width=0.4 \textwidth]{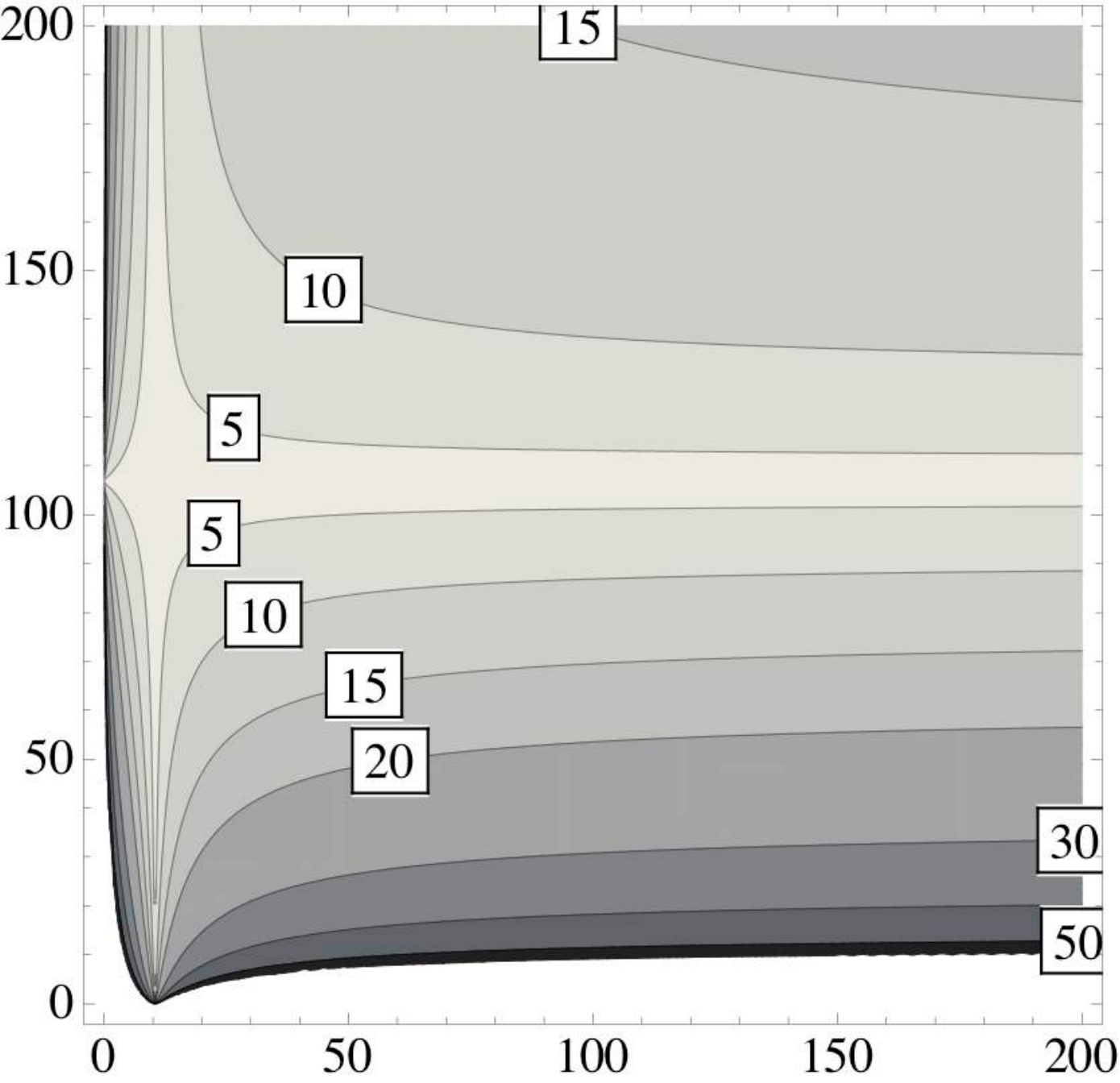}
\put(-82,140){\fcolorbox{black}{white}{$\begin{array}{c}\kappa_{q_2}= 300 \\ \kappa_{d_2}=1500 \end{array}$} }
\put(-90,-10){$\kappa_{q_1}$}
\put(-200,100){$\kappa_{d_1}$}
\caption{\label{fig:flvbnds} Bounds on KK-scale in the presence of brane localized kinetic terms for the fermions. The strengths of two of these are kept fixed, while we scan over the other two. Left: for $\kappa_{q_1}=\kappa_{d_1}=500$, right: for $\kappa_{q_2}=300$ and $\kappa_{d_2}=1500$}
\end{figure}
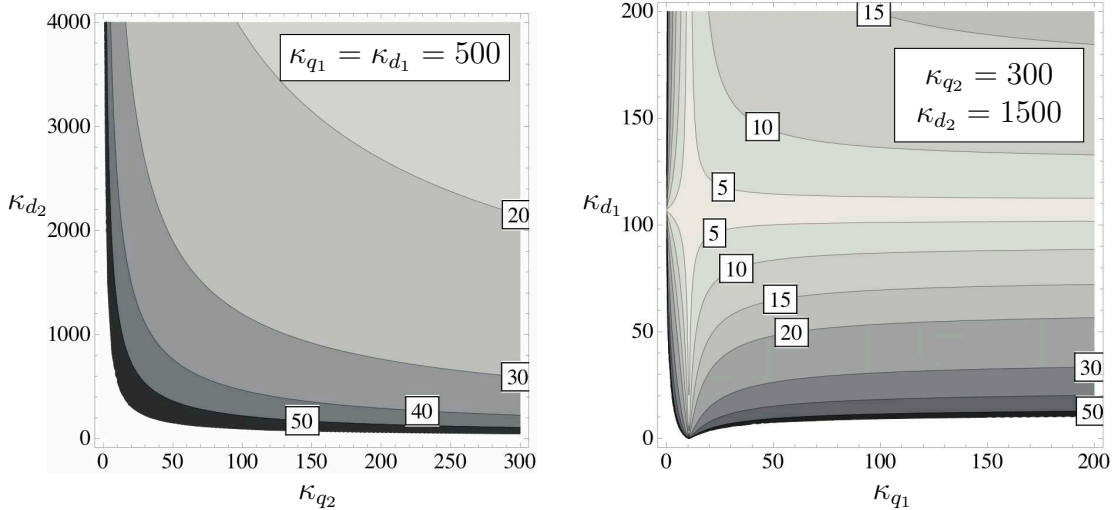

As before we derive the coupling of the second KK-gluon to the SM fermions. As before, we obtain
\begin{equation} \begin{split}
g_s^{(2)}(c, \kappa) 	
		& = g^{4D} \sqrt{2}  \left[1- f_{c,\kappa}^2 \, \gamma^{(2n)}_c \right] ,
\end{split}\label{eq:g's} \end{equation} 
where $\gamma^{(2)}_c$ is the same as given in eq.~\eqref{eq:gcoupling}.
It is now a matter of feeding this relation into eq.~\eqref{eq:C4} in order to obtain the flavor bounds in the presence of these kinetic terms.  
In Figure~\ref{fig:flvbnds}, we plot the dependence of the allowed KK-scale as a function of the different $\kappa$'s that may be added for the different fermions. It can be seen that the flavor bounds are greatly improved for larger values of $\kappa$. 

This is a result of the fact that
including  very large $\kappa_f \sim v/m_f$ mimics the RS GIM mechanism:
The brane kinetic term contributes a universal part to the couplings of gauge KK-modes to the fermion, and simultaneously suppresses the non-universal contribution by diluting the fermion wave function in the remainder of the bulk.\footnote{The $\kappa$'s cannot be too large, however, since a very large $\kappa$ would force a fermion to be localized towards the end-points of the extra dimension, reintroducing the ultra-light mode discussed in Section~\ref{sec:ultralight}.}

In RS, the gauge boson KK-modes are basically flat throughout most of the bulk of the extra dimension, varying mainly in the region of the IR brane.  Integrating out the region in the vicinity of the UV brane then creates (after canonical normalization of the zero modes) flavor universal brane localized gauge-covariant kinetic terms.  The remaining non-universal pieces arise only near the IR brane, where the fermion wave functions are exponentially suppressed.  This is the essence of the natural RS GIM, which we have just shown can be forced in a rather unnatural way into a flat extra dimension to obtain the same effect.

This behavior can also be seen  more explicitly:
 For large $\kappa$, a smaller $|c|$ is required to obtain a light fermion mass, 
which leads to a smaller value of $\gamma^{(2)}_c$. For $\gamma^{(2)}_cÊ\sim \mathcal{O}(1)$ we see from eq.~\eqref{eq:offdiagcoupling} that the off diagonal couplings are given by
\begin{equation}
g_{ij} \sim \sqrt{2} g_s  f_i f_j.
\end{equation}
This results in an effective suppression of the KK-gluon FCNC contribution
\begin{equation}
C_K^4 \sim \frac{2 g_s^2}{(M_G^{(2)})^2} f_{q_1} f_{q_2} f_{d_1} f_{d_2}.
\end{equation}
As anticipated, this is the same relation that is obtained in RS models~\cite{Csaki:2008zd}.

We note that we have been forced to reintroduce the same flavor problem present in original UED, with flat wave-functions.  In the absence of an underlying flavor symmetry, the gauge-covariant brane kinetic terms that we add at $y=0$ can be misaligned in flavor space.   If such terms are also added on the endpoints, with similar magnitude, it will create disastrously large contributions to FCNC's.   This is a key conclusion of this paper; UED does not seem to admit simultaneously a natural mechanism for generating the fermion mass hierarchy while avoiding bounds on observables sensitive to flavor.

%%%%%%%%%%%%%%%%%%%%%%%%%%%%%%%%%%%%%%%%%%%%%%%%%%%%%%%%%%%%
%%%%%%%%%%%%%%%%%%%%%%%%%%%%%%%%%%%%%%%%%%%%%%%%%%%%%%%%%%%
\section{Other indirect constraints} \label{sec:EWPC}

%%%%%%%%%%%%%%%%%%%%%%%%%%%%%%%%%%%%%%%%%%%%%%%%%%%%%%%%%%
\subsection{Electroweak precision bounds}

In the original UED model, a flat Higgs VEV leads to vanishing contributions to electroweak precision observables ($S$, $T$, and $U$ parameters) at tree-level. Our setup, however, employs a Higgs that is localized at the two boundaries of the extra dimension. This will lead to sizable corrections  that constraint our model.

We estimate the size of shifts in electroweak precision  observables
due to the localization of the Higgs at the boundaries. For simplicity we assume that the Higgs is completely localized on branes at $y=\pm L/2$.
The terms in the 5D Lagrangian relevant to EWP are:
\begin{equation}
\int dz \frac{g^2 v_0^2}{8} \left[ W^{(1)2}_\mu + W^{(3)2}_\mu - 2 \frac{g'}{g} W^{(3)}_\mu B^\mu \right] \frac{\delta(y+L/2)+\delta(y-L/2)}{2} 
\end{equation}
After expanding this in terms of the KK-modes,  we keep only terms  which yield a mass mixing between the SM mode and the
higher KK-modes
\begin{equation}
\sum_n \int dz \frac{g^2  v^2_0}{4} \left[ W^{(1)}_{0 \mu}
W^{(1)\mu}_n + W^{(3)}_{0 \mu} W^{(3)\mu}_n - \frac{g'}{g}
W^{(3)}_{0\mu} B^\mu_n \right]  \frac{\delta(y+L/2)+\delta(y-L/2)}{2}.
\label{eq:ewplag}\end{equation} 
The diagrams involving heavy charged $W$ exchange cancel in  calculating
$\Pi_{11} - \Pi_{33}$, so we need only calculate the diagrams mixing
the heavy $B$ with $W^{(3)}_0$ the last term in
eq.~(\ref{eq:ewplag}).
The  overlap integrals for the mixing terms are given by
\begin{equation}
\frac{g g' v^2_0}{4 \sqrt{2} L} \times \left[Ê1 + (-1)^n\right]  =\frac{g g'  v_0^2 }{2\sqrt{2}L}
\cdot \left\{ \begin{array}{ll} 1 & n~\mathrm{even} \\ 0 &
n~\mathrm{odd} \end{array} \right.
\end{equation}
Therefore, the diagrams contributing to a shift in electroweak precision observables  evaluate to:
\begin{equation}
g^2 \left( \Pi_{11} - \Pi_{33} \right) =  \sum_{n~\mathrm{even}}
\frac{g^2 g'^2 L^2 v_0^4 }{8 n^2 \pi^2} 
=   \frac{g^2 g'^2 L^2 v_0^4 }{192}
\end{equation}
where  we have used the fact that the masses of the hypercharge
gauge boson KK-modes are approximately given by $m_n = \frac{n
\pi}{L}$. $\Delta \rho$ is then given by:
\begin{equation}
\Delta \rho =  \alpha T = \frac{4}{v_0^2} \left( \Pi_{11} - \Pi_{33}
\right) = \frac{g'^2 L^2 v_0^2}{48} \approx 0.1 \cdot 10^{-3} \times \left( \frac{L}{1 \text{ TeV}^{-1}} \right)^2.
\end{equation}
Depending on the exact value of the compactification scale $L$, this can be in agreement with  the EWP bound of $\Delta \rho \approx (0.1 \pm 0.6) \cdot 10^{-3}$~\cite{Barbieri:2004qk,Alcaraz:2006mx}.  A more detailed study including a full electroweak fit is necessary to find the true limits on the size of the extra dimension, but is beyond the scope of this paper.   We do, however, propose a solution to protect this model against EWP bounds in appendix~\ref{sec:newsetup}: By gauging an extra $U(1)$ symmetry in the bulk Yukawa couplings are only allowed on the boundaries. This allows one to obtain fermion hierarchies (and flavor bounds) as we did in section~\ref{sec:flavor}, while at the same time maintaining a flat Higgs VEV profile.  With a flat Higgs profile, the contributions to electroweak precision are vanishing at tree level, as in original 5D UED~\cite{Appelquist:2000nn}

%%%%%%%%%%%%%%%%%%%%%%%%%%%%%%%%%%%%%%%%%%%%%%%%%%%%%%%%%%
\subsection{Bounds from four-fermion contact interactions}

In the case that flavor and EWP  bounds are satisfied, there are still tree-level contributions to 4-fermi operators which are tightly constrained~\cite{Eichten:1983hw,Barger:1997nf,Cheung:2001wx}.  In the original UED model, these are vanishing at tree-level due to KK-number conservation, but this is not the case in our construction, in which the SM fermions are localized on a domain wall in the center of the extra dimension.

The most stringent bounds are coming from electron-quark contact interactions, parameterized by the following effective Lagrangian,  
\begin{eqnarray}
\mathcal{L}_{\text{eff}} = \frac{4 \pi}{\Lambda^2} \sum_{i,j=L,R} \eta_{ij} \bar{e}_i \gamma_\mu e_i \bar{q}_j \gamma^\mu q_j \, ,
\end{eqnarray}
where $e_i$ and $q_i $ are left or right-handed Weyl spinors.  

The extra-dimensional excitations which contribute most strongly to such operators are the level-2 $SU(2)$ and $U(1)$ gauge bosons (it is sufficient to work in the electroweak symmetric phase as the Higgs VEV only produces small mixing between the massive KK states).  The coefficients are not difficult to compute in terms of the electroweak quantum numbers of the SM fermions and the fermion wave functions.  We have already calculated the overall strength of the couplings in the presence of fermion kinetic terms on the domain wall.  The contribution to the $\bar{e}_L \gamma^\mu e_L \bar{d}_L \gamma_\mu d_L$ operator (the one with tightest bounds) is given by:
\begin{equation}
\frac{g_{L e_L}^{(2)} g_{L d_L}^{(2)}}{(M_{\text{KK}}^{(2)} )^2 } \bar{e}_L \gamma^\mu e_L \bar{d}_L \gamma_\mu d_L
\end{equation}
where the coupling constants $g_{L e_L}^{(2)}, g_{L d_L}^{(2)}$ are given in eq.~(\ref{eq:g's}).  In the limit of large $\kappa$'s, the region in parameter space where flavor bounds may be satisfied, these are simply $g_{L e_L}^{(2)}, g_{L d_L}^{(2)} = g^\text{4D} \sqrt{2}$.  Taking into account $SU(2)_L$ and $U(1)_Y$ quantum numbers in the 4D coupling, we find that bounds on such operators constrain $M_\text{KK}^{(2)} \gsim 3.3~\text{TeV}$, or $L \lsim ( 530 ~\text{GeV} )^{-1}$.

%%%%%%%%%%%%%%%%%%%%%%%%%%%%%%%%%%%%%%%%%%%%%%%%%%%%%%%%%%%%
%%%%%%%%%%%%%%%%%%%%%%%%%%%%%%%%%%%%%%%%%%%%%%%%%%%%%%%%%%%%
\section{Conclusions} \label{sec:conclusion}

We have explored the potential of a 5D flat geometry model with all SM fields propagating in the bulk for generating the observed fermion mass hierarchy while respecting low energy flavor physics bounds and preserving KK-parity.  This is non-trivial, as standard UED is not automatically minimally flavor violating, as has been claimed in earlier literature.  While it is not very difficult to generate the fermion mass hierarchy utilizing wave-function localization in the extra dimension, the resulting tension from flavor physics bounds renders the model rather implausible.  The effects of the Randall-Sundrum GIM mechanism can be mimicked in this construction, but at the price of reintroducing what is essentially the same tuning that is required in the original UED models.  While UED remains an interesting ``straw man" from the perspective of model discrimination at the LHC, it has so far resisted implementation as a theoretically well-motivated competitor with TeV scale supersymmetry, strongly coupled theories of electroweak symmetry breaking, or warped extra dimensions.

%%%%%%%%%%%%%%%%%%%%%%%%%%%%%%%%%%%%%%%%%%%%%%%%%%%%%%%%%%%%
%%%%%%%%%%%%%%%%%%%%%%%%%%%%%%%%%%%%%%%%%%%%%%%%%%%%%%%%%%%%
\section{Acknowledgements}
SCP would like to thank Kingman Cheung for discussion on bounds from four-fermi contact interactions.  J.He.~thanks the IPMU for their hospitality during some course of this research.  J.Hu.~thanks Cornell University for hospitality during the course of this research.  The work of C.C.~and J.He.~is supported in part by the NSF under grant
PHY-0355005.  J.Hu. is supported in part by the DOE under grant number DE-FG02-85ER40237, and in part by the Syracuse University College of Arts and Sciences.  SCP and JS are supported by the World Premier International Research Center Initiative (WPI initiative) by MEXT, Japan and also by the Grant-in-Aid for scientific research (Young Scientists (B) 21740172) and (Young Scientists (B) 21740169) from JSPS, respectively. 
%%%%%%%%%%%%%%%%%%%%%%%%%%%%%%%%%%%%%%%%%%%%%%%%%%%%%%%%%%%%
%%%%%%%%%%%%%%%%%%%%%%%%%%%%%%%%%%%%%%%%%%%%%%%%%%%%%%%%%%%%
\begin{appendix}

\section{The fermion mass specturm} \label{app:fermions}

In this appendix we want to  show the details of calculation for the fermion mass spectrum.
We have a bulk 5D fermion with a varying bulk mass.  The geometry of the extra dimension is flat, and the extra dimensional coordinate, $y$, varies between two endpoints, $y=\pm L/2$:
\begin{equation}
        \label{eq:BulkActionapp}
 S = \int d^4 x \int_{-L/2}^{+L/2} dy
\left[\frac{i}{2}
\bar{\Psi}\, \Gamma^M \overleftrightarrow{\partial}_M \Psi
 - m_5(y) \bar{\Psi} \Psi  \right],
\end{equation}
where $\bar\Psi \overleftrightarrow{\partial_M} \Psi = \bar\Psi \partial_M \Psi - (\partial_M \bar\Psi) \Psi$ and the  gamma matrices are $\Gamma^M= (\gamma^\mu, i\gamma^5)$.
We introduce a $y$-dependent fermion mass which is a step-function that changes sign at the midpoint of the extra-dimension:
\begin{eqnarray}
m_5(y) = \mu ~\epsilon(y) = \left\{ \begin{array}{rl} -\mu, & y<0 \\Ê+\mu, & y >0Ê\end{array} \right.
\end{eqnarray}
Due to  the \emph{intrisic} KK-parity of the 5D Dirac fermion compensates this mass terms is indeed KK-parity in variant unlike a constant (or in general, even under KK-parity) fermion bulk mass~\cite{}.  
After decomposing the Dirac fermion into its left- and right-handed chiral components, 
\begin{equation}
\Psi = \left( \begin{array}{c} \chi_\alpha \\ \bar{\psi}^{\dot{\alpha}} \end{array}\right) 
\end{equation}
we obtain the equations of motion
\begin{eqnarray}
&& -i \bar{\sigma}^\mu \partial_\mu \chi + \partial_5 \bar{\psi} + m \bar{\psi} =0 \nonumber \\
&& -i \sigma^\mu \partial_\mu \bar{\psi} - \partial_5 \chi + m \chi = 0
\end{eqnarray}
%
%%%%%%%%%%%%%%%%%%%%%%%%%%%%%%%%%%%%%%%%%%%%%%%%%%%%%%%%%%%%
%\subsection{KK decomposition and wave equations}
%
Plugging the Kaluza--Klein decomposition of these fields into the equations of motion
 we get the following set of coupled first order differential equations for fermion wave functions:
\begin{equation} \label{eq:1st}
\Psi_n = \left( \begin{array}{c} g_n(y) \chi_{n} (x) \\ f_n (y) \bar{\psi}_n(x) \end{array} \right):~~~~~~~~\begin{array}{l}  - m_n g_n + \partial_5 f_n + m f_n = 0 \\ -m_n f_n -\partial_5 g_n + m g_n = 0 \end{array}
\end{equation}
We obtain a chiral 4D spectrum by projecting out one chirality using boundary conditions: Either $g_{n}|_{y= \pm L/2} = 0$ to obtain a right-handed zero mode, or $f_{n} |_{y=\pm L/2}=0$ to get a left-handed zero mode.  Imposing the bulk equations of motion on the wave functions, eq.~\eqref{eq:1st}, then gives us the boundary condition for the other chirality. 

%%%%%%%%%%%%%%%%%%%%%%%%%%%%%%
\subsection{Zero Modes}

For $n=0$, there are massless solutions to these equations of motion, or zero modes, with wave functions
\begin{eqnarray}
 f_0 (y) = \sqrt{\frac{+ \mu}{1-e^{- \mu L}}} e^{- \mu |y|} ,~~~~~~~g_0 (y) = \sqrt{\frac{- \mu}{1-e^{+ \mu L}}} e^{+ \mu |y|} .
\end{eqnarray}
The wave functions of these zero modes are governed by the boundary conditions (which determine whether the zero mode is LH or RS) and the sign of $\mu$.  For example, if $\mu>0$, and the zero mode is RH, $f_0$ is peaked towards the mid-point of the extra dimension ($y=0$).  There are four possible cases, which we display in Table~\ref{tab:zeromodes}.

As we will discuss in further detail below, the localization of the zero mode determines the mass of the first KK-mode.  We note in particular that in each case where there is a zero mode which is ``doubly-localized" (where the wave function is sharply peaked towards each of the end-points of the extra dimension), there is a KK-mode whose mass is highly suppressed compared with the inverse size of the extra dimension.

%%%%%%%%%%%%%%%%%%%%%%%%%%%%%%
\subsection{KK modes}
In the bulk we can decouple the first order equations for the LH and RH wave functions in the usual way, obtaining identical second order equations of motion for the $f_n$'s and $g_n$'s.  
\begin{equation}
\begin{array}{l} (\partial_5^2 +\Delta_n^2)f_n (y) = 0 \\ (\partial_5^2 +\Delta_n^2)g_n (y) = 0 \end{array}, \qquad \text{ with }  \Delta_n^2 \equiv m_n^2 -m_5^2
\end{equation}
Depending on the sign of $\Delta_n^2$ the solutions will be either sines and cosines or sinhes and coshes.  In principle, the bulk second order equation of motion contains a delta function contribution from the $y$-derivative acting on the bulk mass term.  We take this into account by separately solving the equations of motion in the two regions to the left and right of the discontinuity in the bulk mass, and matching the solutions at the discontinuity appropriately.

\begin{center} $\mathbf{\Delta_n^2 = k_n^2 >0}$ \end{center}

 The wave functions are given by
\begin{eqnarray} \label{eq:wv1}
&&f_n(y) = A_n \cos k_n y + B_n \sin k_n y \nonumber \\
&&g_n(y) = C_n \cos k_n y + D_n \sin k_n y,
\end{eqnarray} 
Where there are separate coefficients for the wave functions in the regions $y<0$ and for $y>0$.
Imposing the first order bulk equations of motion on these solutions eliminates four of the eight undetermined coefficients.  The rest are eliminated by imposing the boundary conditions, and continuity condition for the wave functions across the jump in bulk mass.  These also determine the mass eigenvalues to be 
\begin{equation} \begin{split}
n=2,4,...& : k_n = \frac{n \pi}{L} \\
n=(1,)3,... & : k_n = \mp \mu \tan \frac{k_n L}{2} \qquad \text{ for RH/LH zero mode} \\
\end{split} \end{equation}
The brackets for the $n=1$ mode are supposed to indicate that this heavy mode with $m_1 > \mu$ does \emph{not 	always} exist, depending on the sign of $\mu$ and the imposed boundary conditions. but can rather be much lighter than $\mu$ (see next paragraphs).

\begin{center} $\mathbf{\Delta_n^2 = - \kappa_n^2 < 0}$ \end{center}

 In this case, the wave functions are given by
	\begin{eqnarray} \label{eq:wv2}
		f^n(y) = \alpha^n \cosh \kappa_n y + \beta^n \sinh \kappa_n y \nonumber \\
		g^n(y) = \gamma^n \cosh \kappa_n y + \delta^n \sinh \kappa_n y 
	\end{eqnarray}
	The boundary conditions and bulk equations of motion allow for at most one mass eigenvalue in this regime, determined by 
	\begin{equation} \begin{split}
		(n=1) & : \kappa_n = \mp \mu \tanh \frac{\kappa_n L}{2} \qquad \text{ for RH/LH zero mode boundary conditions} \\
	\end{split} \end{equation}
	This equation only has solutions for $\mp \mu > 0$. Under this assumption, and if $\mu L \gg 1$, the corresponding mass eigenvalue is then given by\footnote{For large $x$: $\tanh x \approx 1 - 2 e^{-2 x}$} $m_1^2 = \mu^2 - \kappa_1^2 \approx 4 \mu^2 e^{-\mu L}$. For $L \sim 1/\text{TeV} $ these light modes would be problematic and we choose boundary conditions to avoid them in our attempt to build a model of UED flavor.

The dependence of the spectrum on $\mu$ is shown in figure~\ref{Fig:spectrum}, where a RH zero mode is assumed.  For negative $\mu$, the RH zero mode is localized towards the center of the extra dimension, at the kink in the bulk mass.  However, when $\mu$ is positive, this zero mode is localized towards the endpoints.  In this latter case, there is also an ``ultralight" mode whose mass is exponentially suppressed, as derived above.

%%%%%%%%%%%%%%%%%%%%%%%%%%%%%%
\subsection{Ultra-light modes after EWSB} \label{app:ultralightEWSB}

\begin{figure}[t]%[htbp]
\centering
\includegraphics[width=0.65 \textwidth]{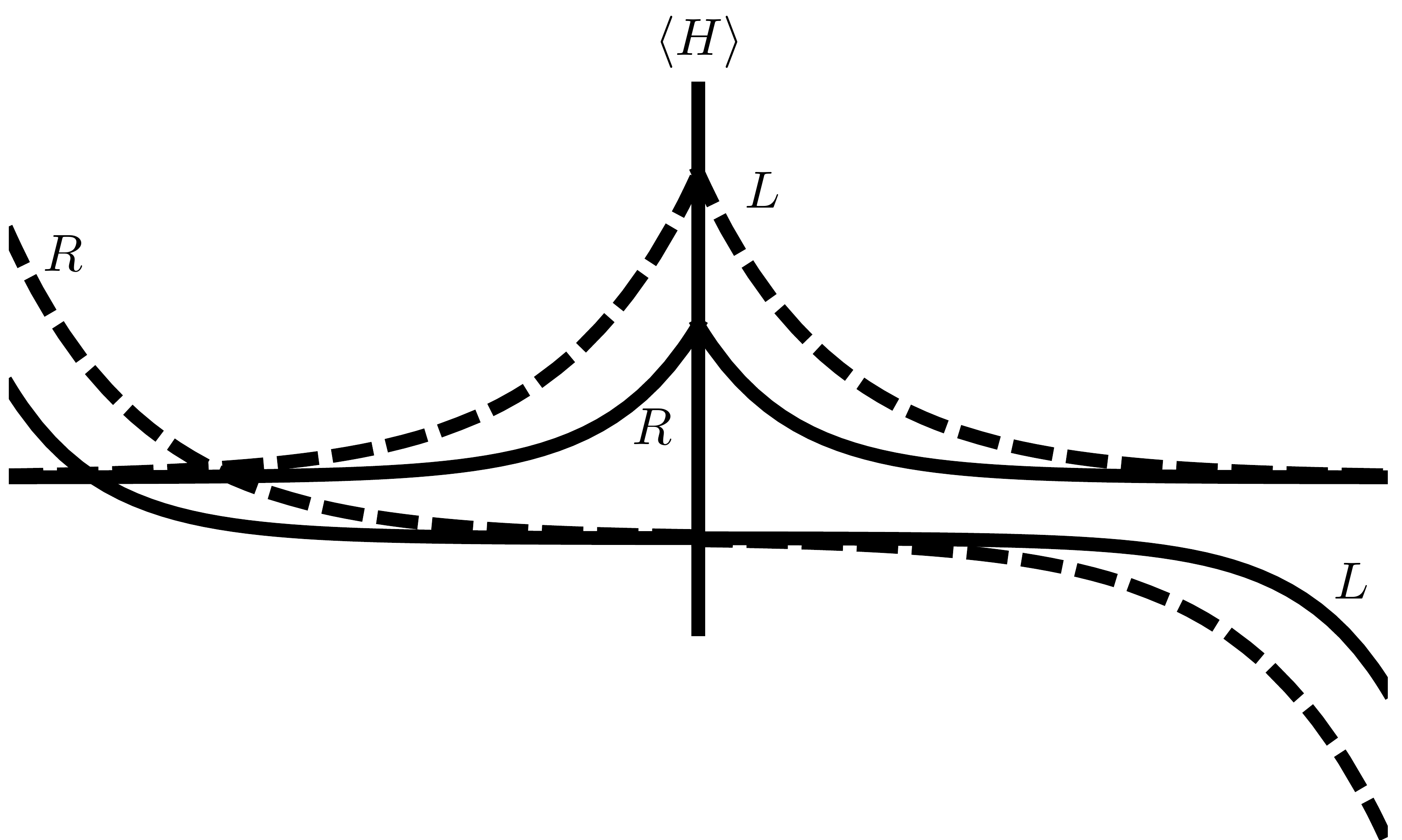}
\caption{\label{fig:lightmodes} We illustrate how the Higgs VEV lifts one ultralight mode from the spectrum, leaving one light mode per SM fermion.}
\end{figure}

The spectrum of light modes is modified in the presence of a Higgs mechanism.  Consider, for example, a model where the Higgs VEV is localized in the center of the extra dimension, on the domain wall.  The SM fermion mass hierarchy could be generated by utilizing doubly-localized zero modes, which have exponentially small overlap with the center-localized Higgs.  For each SM doubly localized zero mode, there is one ultralight Dirac fermion, so for each massive SM fermion, there are two ultralight modes, or four approximately chiral modes.  Two of these have large overlap with the Higgs VEV, and are lifted, while the remaining two have only small overlap with the Higgs VEV, and remain light.  This remaining light is phenomenologically not viable, and we thus prefer to have the SM fermions localized at the center of the extra dimension, avoiding all ultra-light modes.  A diagram showing the pairing of modes is shown in Figure~\ref{fig:lightmodes}.

%%%%%%%%%%%%%%%%%%%%%%%%%%%%%%%%%%%%%%%%%%%%%%%%%%%%%%%%%%%%
%%%%%%%%%%%%%%%%%%%%%%%%%%%%%%%%%%%%%%%%%%%%%%%%%%%%%%%%%%%%
\section{Mass Hierarchies from UED with a flat Higgs profile} \label{sec:newsetup}

It might be desirable to have a model in which the light KK-odd scalar modes are not only lifted by quantum loops, but get a mass already at tree-level.  Additionally, it may be necessary to avoid EWP constraints to have a flat Higgs profile.
The purpose of this appendix is to provide a modified setup that realizes this goal. This modified setup closely resembles that of~\cite{5Dgoldstones}.

%%%%%%%%%%%%%%%%%%%%%%%%%%%%%%%%%%%%%%%%%%%%%%%%%%%%%%%%%%%%
\subsection{Introducing a new gauged $U(1)_X$ in the bulk}
A different way to achieve the fermion mass hierarchy is to find a way in which to forbid the SM Yukawa couplings to the Higgs in the bulk.  This would require gauging a symmetry which forbids Yukawa couplings in the extra dimension, and then breaking it on the endpoints to allow brane-localized Yukawa couplings.  
The easiest possibility is to gauge a $U(1)_X$ symmetry with gauge fields $B_M$ in the bulk.  The complete 5D gauge group is now
\begin{equation}
SU(3)_c \times SU(2)_L \times U(1)_Y \times U(1)_X.
\end{equation}
The Higgs carries charge $Q_X(H)=1$, while the fermions are not charged, and Yukawa couplings are clearly forbidden by the $U(1)_X$ gauge symmetry.

However, the $U(1)_X$ is broken by boundary conditions:
\begin{equation}
B_\mu (z=\pm \tfrac{L}{2}) = 0 \qquad \text{and}Ê\qquad \partial_5 B_5(z=\pm \tfrac{L}{2}) = 0.
\end{equation}
This means that $U(1)_X$ is only a global symmetry on the boundaries, and as such can be explicitly broken by adding Yukawa couplings on the boundary. 
\begin{equation}
{\mathcal L}_Y  = \left[ H \bar \Psi_q Y_u \Psi_{u} + H^* \bar \Psi_q Y_d \Psi_{d} \right]_{y=\pm L/2}  + h.c. 
\end{equation}
The fermion wave functions (which are suppressed on the boundaries by localization) will create the necessary hierarchy, even though the Higgs is not localized at the boundaries.  With a flat Higgs profile, the additional global symmetries discussed in section~\ref{sec:KKoddGoldstone} are strongly broken at tree level, avoiding any extra light KK-odd Goldstone bosons from the Higgs scalars.  An additional attractive feature of this model is that electroweak precision constraints will be reduced, as the $W$ and $Z$ boson wave functions will not be deformed by a flat VEV profile.

However, we have added an extra light degree of freedom, a KK-odd light scalar field, bringing back the problem we sought to avoid.   Taking into account electroweak symmetry breaking, this massless mode is a mixture of the neutral pseudoscalar in the Higgs doublet, and the $B_5$.  The origin of this goldstone boson can be traced to a residual shift symmetry from the original 5D $U(1)_X$ that remains after gauge fixing.

We can get a mass term for this light mode without introducing extra light degrees of freedom by introducing  a two Higgs doublet model (2HDM), with both Higgses charged as $Q_X(H_{u,d})=1$. Yukawa couplings are still forbidden in the bulk and so is the $\mu$-term $H_u^T (i \tau_2) H_d$, but we can add this explicit $U(1)_X$ breaking on the boundaries 
\begin{equation}
\mathcal{L}_{\begin{picture}(20,0)(0,0)
        \put(0,0){\scriptsize $U(1)_X$}
        \put(-2,-2){\line(5,2){24}}  	
        \end{picture}}			= \mu H_u^T (i \tau_2) H_d \big|_{y=\pm L/2},
\end{equation}
which will give the Goldstones a tree level mass. The details of this calculation can be found in~\cite{5Dgoldstones}.

It is also the case that top-quark loops will generate a mass for this mode in the case of just one Higgs doublet.  The $B_5$ mixes with the neutral pseudo-Goldstone, and the Coleman-Weinberg potential generates a mass suppressed by the mixing angle:
\begin{equation}
m^2_\text{odd} \approx \frac{3 \lambda_t^2}{16 \pi^2} M_\text{KK}^2 \left( g_X^2 v^2 L \right),
\end{equation}
where $g_X$ is the 5D gauge coupling, and $v$ is the electroweak VEV.

\end{appendix}

%%%%%%%%%%%%%%%%%%%%%%%%%%%%%%%%%%%%%%%%%%%%%%%%%%%%%%%%%%%%
%%%%%%%%%%%%%%%%%%%%%%%%%%%%%%%%%%%%%%%%%%%%%%%%%%%%%%%%%%%%

\end{document}